\documentclass{aa}
\usepackage{lachaume,txfonts,url}
\usepackage[bf]{subfigure}

\renewcommand{\thesubfigure}{\thefigure.\alph{subfigure}}
\makeatletter
  \renewcommand{\@thesubfigure}{\thesubfigure:\space}
  \renewcommand{\p@subfigure}{}
\makeatother

\newcommand{\Bp}{\ensuremath{B_\text{p}}}
\newcommand{\thetap}{\ensuremath{\theta_\text{p}}}

\def\Tring{\ensuremath{T_\text{r}}}
\def\wring{\ensuremath{h_\text{r}}}

\def\Tin{\ensuremath{T_\text{in}}}
\def\Rin{\ensuremath{R_\text{in}}}
\def\Rmax{\ensuremath{R_\text{out}}}
\def\Tsun{\ensuremath{T_\odot}}

\def\idisk{\ensuremath{i}}
\def\thetadisk{\ensuremath{\theta}}

\def\hen{\mbox{Hen 3-1191}}

\begin{document}


\title{Resolving the B[e] star Hen 3-1191 at 10\,$\mu$m with VLTI/MIDI\footnote{Based on observations made with the Very Large Telescope
Interferometer of the European Southern Observatory.  Programme ID:
073.C-0757}}

\author{
       R. Lachaume\inst{1,2}
  \and Th. Preibisch\inst{2}
  \and Th. Driebe\inst{2}
  \and G. Weigelt\inst{2}
}
\institute{
      Centro de Radioastronom\'\i a y Astrof\'\i sica UNAM, 
      Apartado Postal 3-72 (Xangari), 
      Morelia, Michoac\'an, 
      Mexico C.P. 58089
\and  
      Max-Planck-Institut f\"ur Radioastronomie, 
      Auf dem H\"ugel 69,
      D-53121 Bonn, 
      Germany
}

\offprints{R. Lachaume}
\mail{r.lachaume@astrosmo.unam.mx}
\date{Received 7 August 2006/Accepted 28 March 2007}
\authorrunning{R. Lachaume et al.}
\titlerunning{Resolving B[e] star Hen 3-1191 at 10\,$\mu$m with VLTI/MIDI}

\abstract{%
   We report spatially resolved, spectrally dispersed $N$-band
   observations of the B[e] star Hen 3-1191 with the MIDI instrument of the
   Very Large Telescope Interferometer.  The object is resolved with a 40\,m
   baseline and has an equivalent uniform disc diameter ranging from 24\,mas at
   8\,{\micron} to 36\,mas at 13\,{\micron}.  The MIDI spectrum and
   visibilities show a curvature which can arise from a weak silicate feature
   in which the object appears $\approx 15$\% larger than in the continuum, but
   this could result from a change in the object's geometry within the band.
   
   We then model Hen's 3-1191 spectral energy distribution (.4--60\,\micron) 
   and N-band visibilities.  Because of the unknown nature for the object,
   we use a wide variety of models for objects with IR excesses.  We 
   find the observations to be consistent with a disc featuring an 
   unusually high mass accretion and a large central gap almost void 
   of matter, an excretion disc, and a binary made of two IR sources.  We 
   are unable to find a circumstellar shell model consistent with the data.
 
   At last, we review the different hypotheses concerning the physical nature
   of the star and conclude that it is neither a Be supergiant nor
   a symbiotic star.  However, we could not discriminate between the
   scenario of a young stellar object featuring an unusually
   strong FU~Orionis-like outburst of mass accretion 
   ($4\text{--}250\times10^{-4}\Msun/\yr$) and that of a protoplanetary nebula 
   with an equatorial mass excretion rate $\gtrsim 4 \times 10^{-5}\,\Msun/\yr$.  In
   both cases, taking the additional presence of an envelope or wind into
   account would result in lower mass flows.
   
   \keywords{%
      Stars: emission-line, Be; individual: Hen 3-1191; AGB and post-AGB;
      planetary systems: protoplanetary discs --
      Infrared: stars --
      Accretion, accretion discs -- 
      Technique: interferometric%
   }%
}

\maketitle


\section{Introduction}

\object{Hen 3-1191} ($\alpha = 16^\text{h}27^\text{m}15''$, $\delta =
-48\degr39'27''$, $V = 13.7$), also known as WRAY 15-1484, SS73 56 or VRMF 31,
is a member of the B[e] spectral class as defined by \citet{Allen72, Allen76,
Lamers98}.  It features:
\begin{itemize}
   \item strong Balmer emission lines;
   \item low-excitation permitted emission lines of low-ionisation metals
      in the optical;
   \item forbidden emission lines of [\ion{Fe}{ii}] and [\ion{O}{i}] in
      the optical, an evidence for a diffuse circumstellar (CS) environment;
   \item a strong near or mid-infrared excess hinting towards hot 
      CS dust.
\end{itemize}
A review of the properties of these still mysterious emission-line stars 
was given by \citet{Zickgraf98}.

Though these properties indicate similar physical conditions in the
CS environment of B[e] stars in terms of temperature, density, and
velocity, these stars form a highly heterogeneous group.  As categorised by
\citet{Lamers98}, they comprise various objects : B[e] supergiants
(sgB[e]), young stellar objects (YSOs) of Herbig type (HAeB[e]), compact
planetary nebulae (cPNB[e]), and symbiotic stars (symB[e]).  Concerning
\hen{}, several photometric and spectrometric studies have led to radically
different conclusions: the star is reported as a cPNB[e] by
\citet{Pereira03,Zickgraf03}, a post-AGB object by \citet{DeWinter98}, a young
stellar object (YSO) by \citet{DeWinter94, The94}, or as a possible symbiotic
object \citep[see][]{DeWinter94}.  This uncertainty arises from the absence
of the major discriminating criteria, the bolometric luminosity -- 
\hen{}'s distance is unknown.  So, \citet{Lamers98} fit the star
into the ad-hoc group of unclassified objects, unclB[e]. 

\citet{DeWinter94} started the modelling work on \hen{} and noticed that
its spectral energy distribution (SED) in the range 0.4--10\,$\micron$
could be explained by an early B star photosphere surrounded by dust emitting
at a temperature of 950\,K.  The IRAS measurements from 12 to 100\,$\micron$
are, however, in excess compared to this model and their slope hint towards the
additional presence of a disc or envelope.  \citet{Elia04} successfully used a
radiative transfer simulation of \hen{}, with a star surrounded by a
spherical CS shell.  This model accounts for the SED from 0.4 to
100\,$\micron$ with the constraint of two polar cavities of relatively small
radial extent.  No detailed simulation of the spectroscopic data has been
carried out so far; studies mainly focus on the presence and width of
particular emission lines.  Recently, \citet{Zickgraf03} derived $19 \pm
2$\,km/s FWHMs for [\ion{O}{I}], [\ion{N}{II}], \ion{Fe}{II}, and
[\ion{Fe}{II}] lines, and a H$\alpha$ peak separation of 79\,km/s.
\begin{table*}[t]
   \centering
   \caption{Log of \hen{} MIDI observations.  The projected baseline length
   {\Bp}, the position angle of the base {\thetap}, and the reduced visibility
   at the boundaries and the centre of the $N$ band are given.  The calibrators
   used for each observing night are given in the bottom rows of the table.}
   
  \input tables/lachaume_tab1.texpart
    
   \label{tab:data}
\end{table*}
\begin{figure}[tp]
   \centering
   \includegraphics[width=.9\linewidth]{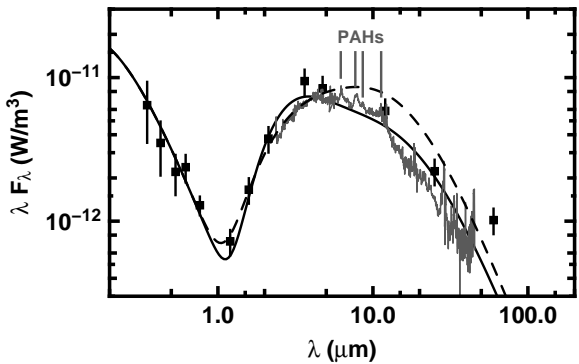}
   \caption{Observed and modelled SED of \hen{}. 
   \emph{Markers:} photometry.
   \emph{Gray line:} ISO SWS spectroscopy.
   \emph{Black solid line:} ``large gap'' model fit
   \emph{Black dashed line:} ``small gap'' model fit 
   } 
   \label{fig:haebe-model-SED}
\end{figure}
\begin{figure*}[t]
   \centering
   \subfigure[Uniform disc diameter]{%
      \includegraphics[width=.48\linewidth]{lachaume_fig2a}
      \label{fig:obs-ud}
   }
   \subfigure[MIDI fluxes vs. ISO \& IRAS]{%
      \includegraphics[width=.48\linewidth]{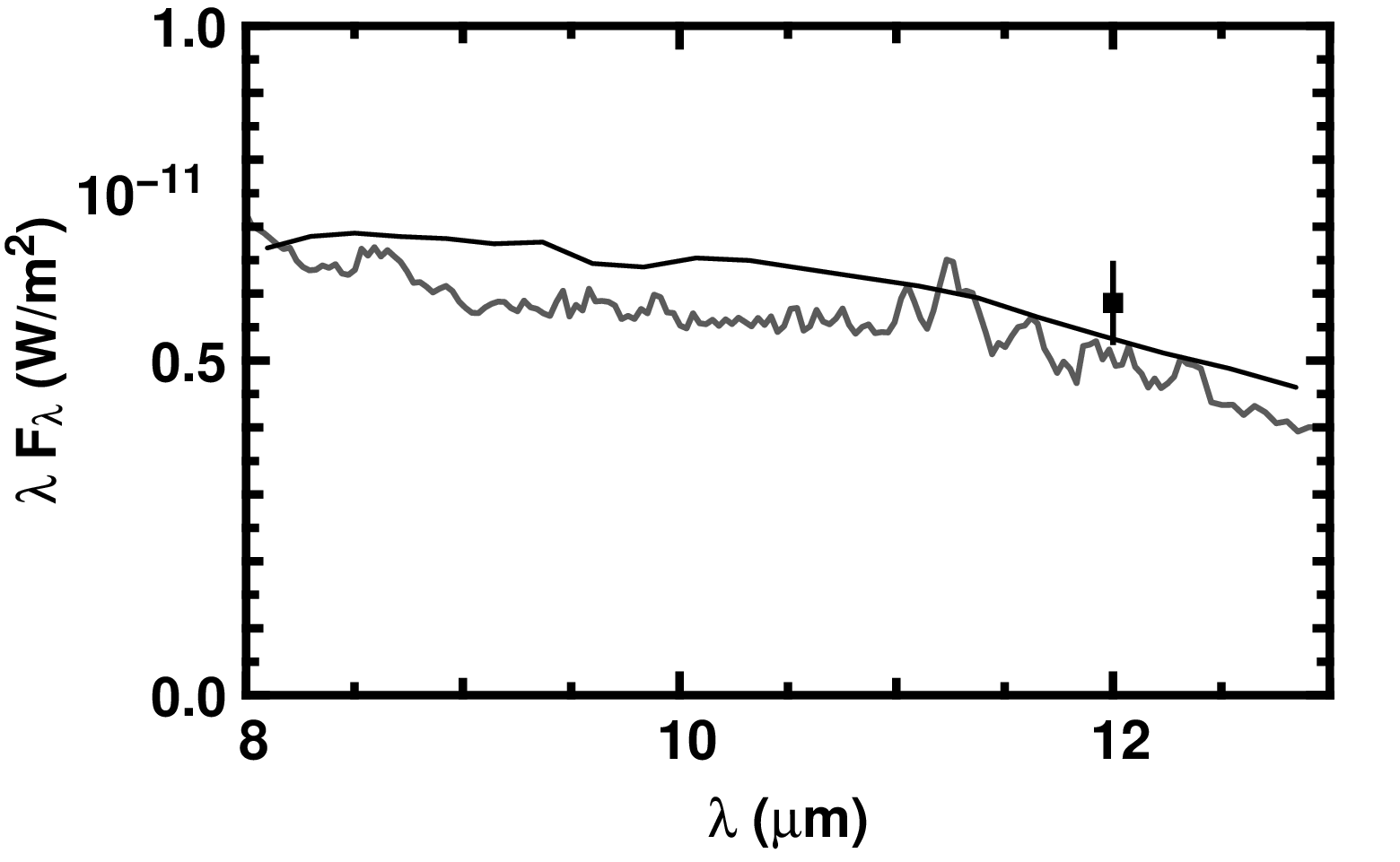}
      \label{fig:obs-midiflux}
   }
   \caption{%
      Observational date of \hen{}.
      \textbf{\ref{fig:obs-midiflux}}
      MIDI fluxes (black solid line) vs. IRAS 12\,\micron{} measurement (square
      with error bar) and ISO SWS flux (gray solid line).  The dashed line
      also represents an estimated continuum value.
      \textbf{\ref{fig:obs-ud}}
      Equivalent uniform disc diameter $\theta_\text{ud}$ (averaged
      over baselines) vs. wavelength.  The markers and error bars represent
      values derived from the observation and the dashed line an estimated 
      continuum value.
   }
\end{figure*}
No X-ray emission has been detected ($L_X < 7.5 \times 10^{-16}$\,W/m$^2$) from 
the star \citep{Hamaguchi05}.

The advent of optical long-baseline interferometry with spatial resolutions of
less than 10\,mas and spectral resolutions up to 10,000 allows us, for the
first time, to geometrically probe the inner parts of stellar environments.
Even with only a few visibility measurements -- giving information on the
mean spatial extent -- it is possible to disentangle SED models and rule out
hypotheses, as shown in recent papers for YSOs \citep{Malbet98, Mil01, Akeson02,
Lachaume03}.  In particular, the Very Large Telescope Interferometer (VLTI)
combined with the $N$ band instrument MIDI gives additional insight in the
regions close to stars. See \citet{Glindemann03} for a recent status
report of the VLTI and \citet{Leinert03} for a description of the MIDI
instrument. 

In this paper, we present spatially resolved and spectrally dispersed
interferometric observations of \hen{} obtained in the $N$ band (8-13\, \micron)
with VLTI (Sect.~\ref{sec:obs}).  In Sect.~\ref{sec:mod} we use the geometrical
information obtained in the previous section and examine three geometrical
models that are encountered in both evolved objects and young stellar objects:
CS shell, CS disc, and binary.  In Sect.~\ref{sec:nat} we
use the models of Sect.~\ref{sec:mod} and review the four hypotheses on
\hen{}'s nature.


\section{Observation and data reduction}
\label{sec:obs}

\subsection{Photometry and spectroscopic data}

The photometric data, presented in Fig.~\ref{fig:haebe-model-SED}, are taken
from various sources: JP11 catalogue in the visible, \citet{DeWinter94} and
2Mass in the near-IR, and IRAS in the far-IR.  The spectroscopy was obtained as
fully reduced and processed data from the Infrared Space Observatory (ISO)
archive. Emission features of polyaromatic hydrocarbons (PAHs) are identified
at $\lambda\lambda$ 6.2, 7.7, 8.6, and 11.3\,{\micron}.  There is no noticeable
silicate feature at 10\,{\micron}.  Visible and near-IR data have been
dereddened using $\Av = 3$ \citep[It allows to fit best the visible data with
an early star, and is the value reported in the model by][]{Elia04}.

\subsection{Interferometric data}

\hen{} was observed with the UT2-UT3 baseline (length 46.6 m, azimuth
40\degr) of the VLTI at Cerro Paranal, Chile.  Three MIDI visibilities with a
spectral resolution of $\approx 25$ in the $N$ band (8--13 \micron) were
obtained at different hour angles, thus, projected baselines. 

The different calibrators observed in the same night predict system
visibilities differing by up to $\approx 0.1$ with no clear connection to
angular distance from the source or to the time of observation.  In order to
minimise the risk of systematical errors coming from one calibrator, the system
visibility was determined using the average of all calibrators observed in one
night. The data reduction was performed with the MIDI Interactive Analysis
reduction software \citep[][incoherent visibility integration in the Fourier
space]{Kohler04}, and we checked that the results were consistent with the
results given by the Expert Work Station software \citep[][coherent visibility
integration in the direct space]{Jaffe04}. The calibrators were also
used to derive the spectrum from the uncorrelated fluxes.

Table~\ref{tab:data} sums up the reduced data obtained at three different
projected baselines (\Bp, \thetap) by giving the calibrated visibility at
the centre of the $N$ band and at its boundaries.  The detailed spectrum of
the visibility is shown in Fig~\ref{fig:haebe-model-V}.

Figure~\ref{fig:obs-ud} displays the uniform disc diameter and their error bars
of Hen 3-1191 derived from all observed visibilities.  At a given wavelength,
the diameter is an average of the ones derived separately for each baseline.
The uncertainty includes the scatter of the different diameters and the
individual uncertainties on these diameters.  The object is clearly resolved
with a equivalent uniform disc diameter increasing from $24 \pm 3$\,mas at
8\,{\micron} to $36 \pm 4$\,mas at 13\,{\micron}.  The increase of diameter is
steeper from 8 to 10\,{\micron} than from 10 to 13\,\micron, which may
arise either from a spectral feature with an excess in size of 10--15\% over
the continuum represented as a solid line in Fig.~\ref{fig:obs-ud}, or from a
modification in the object's geometry over the band.  Not visible in the ISO
SWS spectra, a curvature is also weakly seen in the MIDI spectra (both are
represented in Fig.~\ref{fig:obs-midiflux}).  


\section{Conditions in the circumstellar environment}
\label{sec:mod}

The MIDI visibilities unveil a warm CS environment with a uniform disc diameter
of 24--36\,mas ($\approx 30\,$AU/kpc).  If arising from a flat, wide
silicate feature, the 10\,\micron{} curvature in these observations would hint
towards the presence of micron-sized dust grains \citep[see][for an analysis
around HAe/Be stars]{Boekel05,Kessler06} at this scale.

In this section, we study the three most common scenarios for a stellar object
resolved by means of IR interferometry: a CS disc, a CS envelope, or a binary
system made of unresolved components.

\subsection{The circumstellar disc scenario}
\label{sec:disk}

\begin{figure}[tp]
   \centering
   \includegraphics[width=\linewidth]{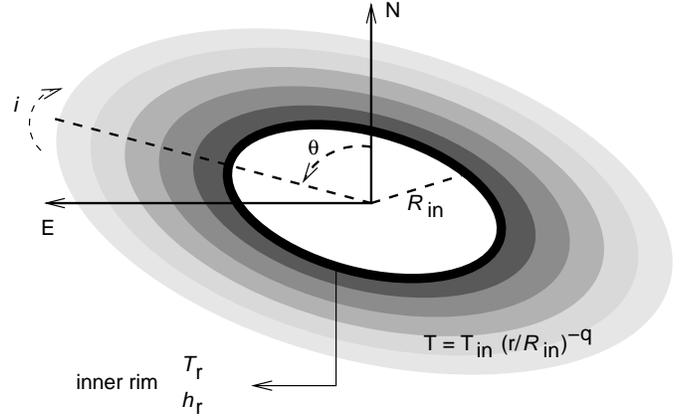}
   \caption{Sketch of an optically thick CS disc around an 
      intermediate-mass star.  It consists of a flat disc with an inner 
      irradiated rim.}
   \label{fig:haebe-model}
\end{figure}

Discs are a common hypothesis around intermediate-mass young stellar objects
\citep{Rieke05} and evolved stars \citep{Porter03}.  In Herbig HAe/Be
stars, the shape of the near-IR excess as well as interferometric observations
give constraints on the geometry: the disc features an inner gap beyond the
dust sublimation radius, a hot, puffed-up inner rim, directly irradiated by 
the star \citep{Natta01,Dullemond01,Monnier05}.  The far-IR excess originates
from the remaining parts of the disc, heated by viscous heating and/or stellar light at
a grazing angle.  A typical sketch of these systems is given by 
\citet[][see their Fig.~1]{Dullemond01}.

\subsubsection{Model}


Our model (see Fig.~\ref{fig:haebe-model}) is a simplification of the 
aforementioned picture.  It uses three flat, infinitely thin elements 
emitting like black bodies:
\begin{itemize}
   \item \emph{Star:} disc with temperature $\Tstar$ and radius $\Rstar$; 
   \item \emph{Inner rim of the disc:} ring of width $\wring$, radius $\Rin$,
     temperature $\Tring$, inclination $\idisk$ (0 if pole-on), and position
     angle $\thetadisk$;
   \item \emph{Remaining parts of the disc:} disc of inner radius 
     $\Rin + \wring/2$, inclination $\idisk$, position angle $\thetadisk$,
     with a radial temperature profile $T(r) = \Tin (r/\Rin)^{-q}$, 
     where $q$ is a dimensionless parameter ranging from $\approx 0.5$ 
     (irradiated disc) to $0.75$ (viscous disc).
\end{itemize}

The method to compute the spectra and visibilities from these geometrical
elements is based on the Hankel transform (disc) or analytical expressions
(star and ring).  Further details are available in \citet{Malbet05}.

In order to obtain a physically relevant model, we self-consistently
determine the location of the inner rim $\Rin$ as a function of
stellar luminosity by assuming that it is located where the dust sublimates.
The ``small gap scenario'' proposed for luminous stars ($\gtrsim 10^3\,\Lsun$)
by \citet{Monnier05}  states that there is enough gas in the sublimation region to shadow the disc 
from the star.  Then, the inner rim is heated by irradiation at a
grazing angle from above the surface of the disc. From Fig. 4 in 
\citet{Monnier05}, we derived
\begin{equation}
   \Rin = 7.0 \times 10^{-2}\,
          \left(\frac\Tstar\Tsun\right)^{1.6}
          \left(\frac\Tring{1000\,\Kelvin}\right)^{-1.2}
          \left(\frac\Rstar{1\,\Rsun}\right)^{0.8}
          \,\AU,
\end{equation}
where $\Tstar$ and $\Rstar$ are the effective temperature and radius of the
star, and $\Tring$ the temperature of the inner rim. The ``large gap
scenario'', as explained by \citet{Dullemond01,Monnier05} states that the rim
is indeed directly heated by the star because the inner region is gas-free or
only contains optically thin gas. Then,
\begin{equation}
   \Rin = 3.0 \times 10^{-3}\,
          (2\epsilon)^{-1/2} 
          \left(\frac\Tstar\Tring\right)^{2} 
          \left(\frac\Rstar{1\,\Rsun}\right)
          \,\AU,
\end{equation}
where $\epsilon$ is the ratio of the ring opacity for its own radiation to
its opacity for the stellar radiation \citep[as in][]{Chiang97}.  

\subsubsection{Results}
\label{sec:disk:res}

\begin{table}[tp]
   \tabcolsep=0.4em
   \centering
   \caption{Best-fit model of \hen{} visibilities and SED with a 
      parametrised disc model.  Figures appear in pairs, the left one
      stands for the ``large gap'' (lg) model, the right one for the ``small
      gap'' (sg) model.
   }
   
  \input tables/lachaume_tab2.texpart
 
   \label{tab:haebe-model-param}
\end{table}

\begin{figure*}[tp]
   \centering
   \includegraphics[width=.9\linewidth]{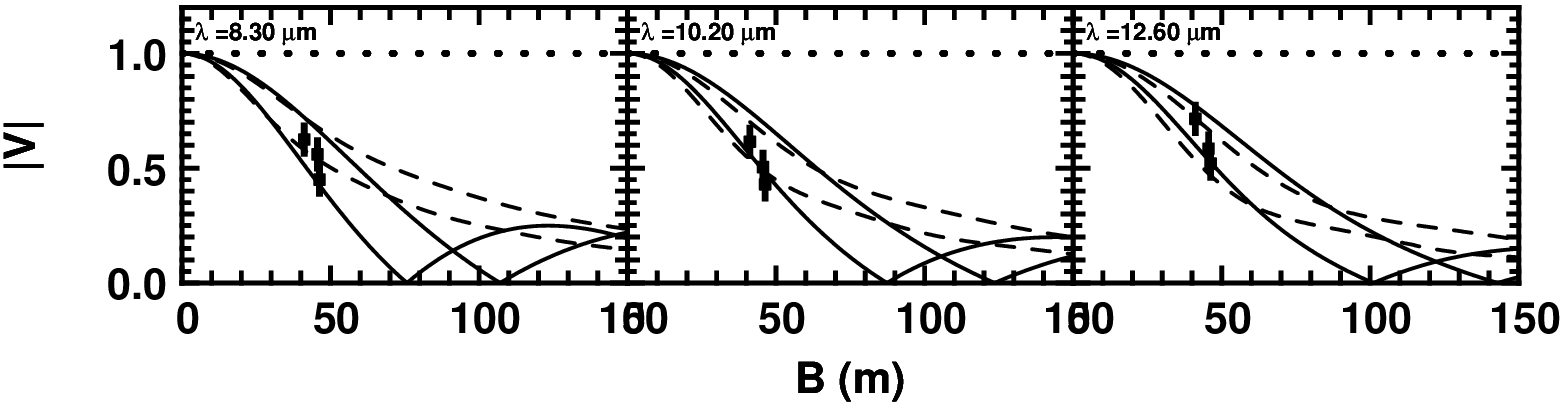}\\
   \includegraphics[width=.9\linewidth]{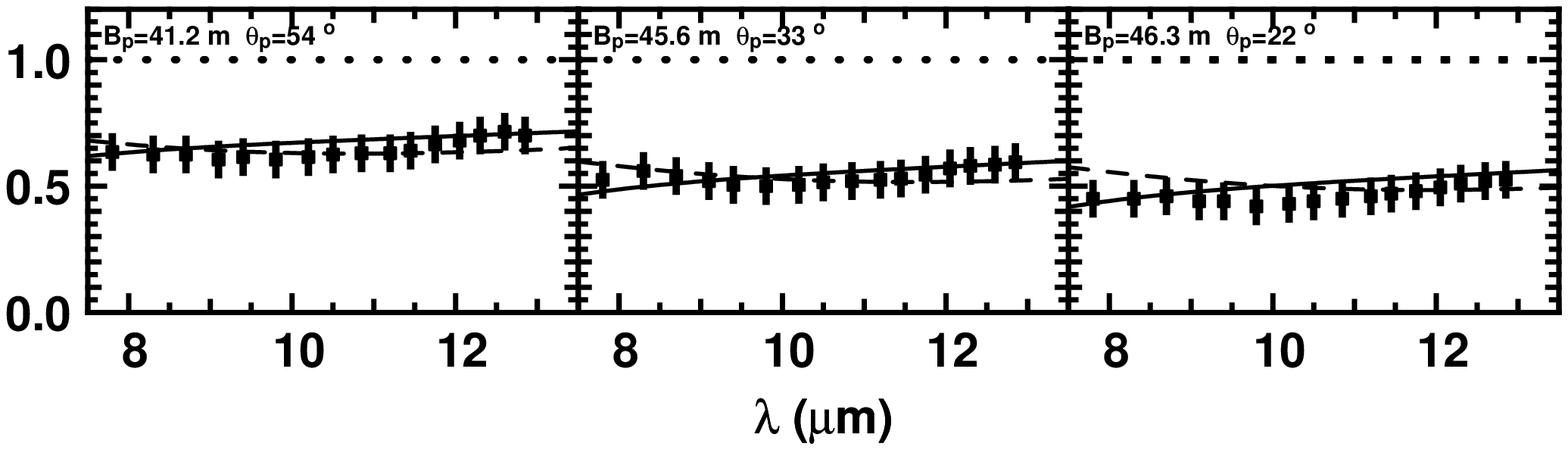}
   \caption{%
     Observed and modelled visibilities of \hen{}.
     \emph{Markers:} MIDI visibilities
     \emph{Black solid line:} ``large gap'' model fit to the SED and visibilities.
     \emph{Black dashed line:} ``small gap'' model fit of the SED and visibilities.
     \emph{Top panel:} visibility vs. baseline at the boundaries and in the
     centre of the $N$ band. The visibility curve along the major axis (lower
     curve of a pair) and the minor axis (upper curve of a pair) of the 
     inclined disc are given.
     \emph{Bottom Panel:} visibility vs. wavelength for each of the
     three baselines.}
   \label{fig:haebe-model-V}
\end{figure*}

We derived best-fit models of the photometric, spectroscopic, and
interferometric data. Due to the large numbers of ISO data (35,000
points) compared to photometry and visibilities (a few tens), a standard
least-square fitting procedure would only give weight to the
spectroscopy.  Therefore, we computed separated reduced chi squares for
the three kind of observables, respectively $\chi^2_\text{ph.}$,
$\chi^2_\text{sp.}$, and $\chi^2_\text{vis}$, and minimised their
average $\chi^2_\text{tot}$.  Combined with the presence of large, 
systematic calibration errors that the standard least-squares method 
does not handle, this algorithm prevents us to obtain reliable 
uncertainties on derived parameters; our estimates were
derived from the geometrical mean of the uncertainties given by the 
photometric and interferometric chi squares.

We achieved a satisfactory agreement between the ``large gap''
model and data, but we were not able to do so for the ``small gap''
hypothesis.  In the latter case, we only minimised using
$\chi^2_\text{ph.}$ and $\chi^2_\text{vis}$; the resulting fit is
clearly not consistent with the spectroscopy ($\chi^2_\text{sp.} =
12.3$).  The model parameters and chi squares are given in
Table~\ref{tab:haebe-model-param}.  The comparison between model and
data are displayed in Figs.~\ref{fig:haebe-model-SED}~\&
\ref{fig:haebe-model-V}.

First, the ``large gap'' model fits the data better than the ``small gap''
model -- only marginally for visibilities, but substantially for
photometry and spectroscopy.  Second, the small gap models leads to physically
inconsistent parameters: the temperature law features $q = 0.50$ hinting
towards a flared, irradiated disc while the inner temperature of the disc at
the inner rim $\Tin \approx 1640\,\Kelvin$ is much too large for
irradiation.  Since the small gap model implies that matter beyond radius
$\Rin$ is irradiated at a grazing angle like the remaining parts of the disc, we
should have $\Tin = \Tring \approx 1000\,\Kelvin$, which is a factor
of 7.5 below in terms of flux.  We therefore conclude that, provided that the
disc scenario is correct, \hen{} features a wide gap (30\,\AU) almost free of
matter or consisting of optically thin gas.

The wide gap models hints towards an active disc (temperature law with $q =
0.76$), which is valid for both an accretion disc \citep{Shakura73} or an
excretion disc \citep{Lee91}. From the disc temperature ($\Tin = 510\,\Kelvin$)
at the inner rim (30\,\AU) we can derive the accretion or excretion rate
$\Mdot$ from the standard laws for each of these discs.  Substituting
$R = \Rin \gg \Rstar $, $R_0 = \Rstar$, and $Q = \sigma\Tin^4$ in 
the first and fourth terms of Eq.~(2.6) by \citet{Shakura73} we obtain the 
accretion  rate
\begin{subequations}
\begin{align}
  \Mdot &= \frac{8\pi\sigma\Tin^4\Rin^3}{3G\Mstar}\\
        &\approx 2.5 \times 10^{-2} 
                     \left(\frac\Mstar{14\,\Msun}\right)^{-1}
                     \left(\frac d{4\,\kpc}\right)^3
                     \ \Msun/\yr
\end{align}
\end{subequations}
where $d$ is the distance to the star if we apply a distance scaling to
$\Rin$ in our model computed for a distance of 4\,kpc. For 
an excretion disc, we carry out the substitutions $\Omega^2 = 
G\Mstar/\Rin^3$, $Q^+_\text{vis} = 2\sigma\Tin^4$, $r = \Rin$, $r_1 = \Rmax 
\gg \Rin$ in Eq.~(12) by \citet{Lee91} and find
\begin{subequations}
\begin{align}
  \Mdot &= \frac{8\pi\sigma\Tin^4\Rin^{7/2}}{3G\Mstar\Rmax^{1/2}}\\
        &\approx 4.3 \times 10^{-3}
                    \left(\frac\Mstar{14\,\Msun}\right)^{-1}
                    \left(\frac\Rmax{1000\,\AU}\right)^{-1/2}
                    \left(\frac d{4\,\kpc}\right)^{7/2}
                \ \Msun/\yr,
\end{align}
\end{subequations}
where $\Rmax$ is the outer radius of the excretion disc.

\subsection{The circumstellar shell scenario}

\subsubsection{Model}

\begin{figure}
   \centering
   \includegraphics{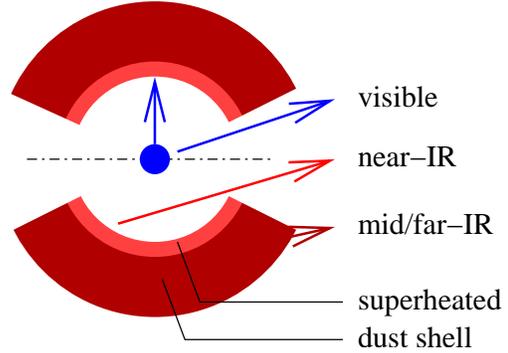}
   \caption{Sketch of the CS shell.  Arrows represent light
   beams.}
   \label{fig:ppn-model}
\end{figure}

\citet{Elia04} successfully explained the SED of \hen{} with a CS shell without
polar holes --- but polar cavities of small radial extension.  Since the star
features a large-scale bipolar nebula \citep{DeWinter94}, probably resulting
from an outflow, we conversely assume the shell should feature polar holes.
The sketch of such a model is represented in Fig.~\ref{fig:ppn-model}.
Furthermore, the star is likely to be seen through this hole, since the
moderate extinction \citep[$\Av \lesssim 3$, see][Table 2]{Elia04} is probably
interstellar, placing the star at the distance of 4-5\,kpc (assuming an early B
main-sequence star) in the galactic plane.

We used the iterative numerical radiative transfer simulation described in
\citet{Sonnhalter95} in order to derive the temperature distribution around the
star from a fixed, assumed density distribution.  Once convergence of radiation
field and temperature are achieved, ray-tracing is performed to obtain images
and fluxes from the density and temperature distribution.  The space within the
self-consistently determined sublimation radius is assumed free of matter, as
are the conic polar holes.  The simulation uses the flux-limited diffusion 
approximation \citep{Levermore81} in a set of concentric square meshes. 
The 2D distribution in the $(r, \theta)$ space is written
\begin{equation}
   \rho(r, \theta) =
   \begin{cases}
      \phantom{\epsilon}  \rho_0  (r/r_0)^{-q}    &\text{if $r > r_0$ and $\theta < \theta_0$}\\
      \epsilon \rho_0 (r/r_0)^{-q}                &\text{if $r > r_0$ and $\theta > \theta_0$}\\
      0                                           &\text{if $r < r_0$}
   \end{cases}
\end{equation}
where $\theta_0$ is the opening angle of the polar holes, $\rho_0$ and $r_0$
the density and location of the sublimation radius, $q$ the density exponent,
and $\epsilon = 10^{-4}$ a small factor used to prevent numerical 
instabilities.

\subsubsection{Results}

\begin{table}
  \centering
  \caption{Parameters of the envelope model.  Instead of the density
  at the inner rim used in the model, the optical thickness in the visible
  is given.}
  \label{tab:env:res}
  
  \input tables/lachaume_tab3.texpart

\end{table}

\begin{figure*}
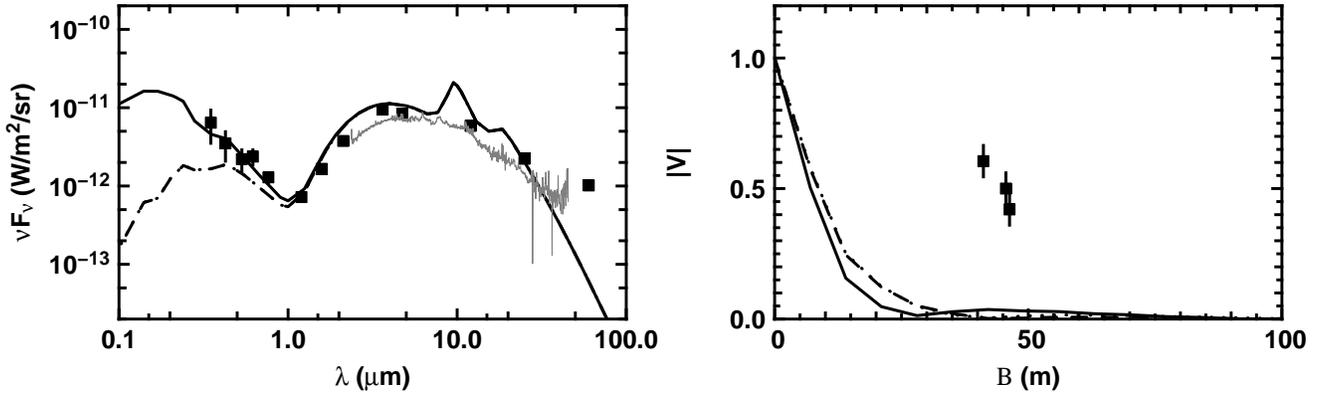

   \centering
   \includegraphics[width=.48\linewidth]{lachaume_fig6a}
   \includegraphics[width=.48\linewidth]{lachaume_fig6b}
   \caption{SED and visibilities of \hen{}: 
   model fit using the spherical shell model with polar holes.  
   \emph{Solid line:} pole-on model. \emph{Dashed line:} equator-on model.
   \emph{Left:} SED.  \emph{Right:} visibilities vs. baseline}
   \label{fig:ppn-sed}
\end{figure*}

We tried to fit SED and visibilities using the envelope model but we failed to
find a model consistent with the observations.  Table~\ref{tab:env:res} lists
the parameters of an envelope model that fits the SED (see
Fig.~\ref{fig:ppn-sed}, left panel), but all models give too large
structures (by a factor of at least two) in order to explain the MIDI
observations (e.g. same figure, right panel).  The optical thickness we derive
from the SED is different from the one found by \citet[][$\tau_V =
2.7$]{Elia04} but the density distribution follows the same power law ($q =
0.5$).  The discrepancy may arise from different opacity laws in our model
($\kappa_\lambda \propto \text{2}$ in the sub-millimeter) and theirs
($\kappa_\lambda \propto \lambda^{-1.2}$), or from the difference in the CS
geometry (our models involves a sublimation cavity free of matter while these
authors assume it contains gas).

We can explain the difficulty to find a consistent fit of visibilities
with an envelope by comparing it with a disc: the envelope occupies a larger
solid angle than the inner disc rim, so that the envelope has to be optically
thin in order to reprocess the same amount of stellar light and feature a
similar IR excess.  Then, it means that light is absorbed further out in the
envelope and that the corresponding uniform disc diameter is larger.

\subsection{The binary scenario}

Since the mid-IR flux is in excess by a factor of 100 compared to that of the
stellar photosphere, a binary system that accounts for the $N$-band
visibilities must consist of two objects featuring a large IR flux, thus a
large amount of CS dust.  Such a picture is complex, for it involves possibly
resolved components, and the few geometrical data available could not constrain
such a model.  We therefore consider a simplistic approach using two unresolved
components; however, we can expect that the flux ratio strongly varies with
wavelength since the SED of IR sources exhibit a large variety of behaviours:
for instance, a companion featuring a flat SED around a companion with a
standard accretion disc has a flux ratio going as $F_\lambda \propto
\lambda^{5/3}$ \citep[][$F_\nu \propto \nu^{1/3}$ in their disc
model]{Shakura73}.  

We decided not to try to explain the SED with this model, since it would
require detailed knowledge of the components.  Instead, we use an ad hoc binary
model made of two point-like sources with linear separation $r$, position angle
$\theta$, and wavelength-dependent flux ratio
\begin{equation}
  f = f_0 \left( \frac{\lambda}{\lambda_0} \right) ^ s,
\end{equation}
where $f_0$ is the flux ratio at the reference wavelength $\lambda_0$, and
$s$ an exponent.

\begin{table}
  \centering
  \caption{Parameters of a binary model fit to the MIDI visibilities.}
  \label{tab:bin:res}
  
  \input tables/lachaume_tab4.texpart

\end{table}

\begin{figure*}
  \centering
  \includegraphics[width=\linewidth]{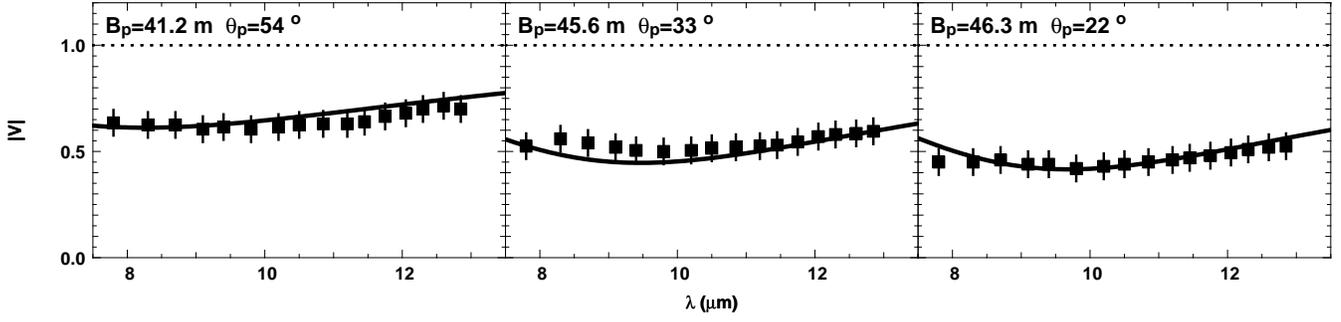}
  \caption{Binary model fit: visibility vs. wavelength at each spatial 
  frequency probed by MIDI.  Line: model; markers with error bars: MIDI data.}
  \label{fig:bin:vis}
\end{figure*}
We found several binary models that fit the MIDI visibilities well, so that the
binary scenario cannot be a priori excluded.  We present the model fits with
the least separation, around 18\,AU per kpc of distance.  Its parameters are 
given in Table~\ref{tab:bin:res} and the fit of MIDI visibilities is 
displayed in Fig.~\ref{fig:bin:vis}.


\section{Nature of the object}
\label{sec:nat}

\subsection{The supergiant hypothesis}

B[e] supergiants with an effective temperature hotter than 20,000\,K -- that of
\hen{} is around 30,000 K -- have a bolometric luminosity greater than
$10^{5}\,\Lsun$ \citep[][see their Fig.  1]{Lamers98}, that is an absolute
bolometric magnitude brighter than $-7.8$. For a B0-B1 star, the bolometric
correction is as small as $-2.8$ \citep[see the catalogue by][]{Bessell98}, so
the absolute visual magnitude is brighter than $-5.0$.  Having a dereddened
magnitude $V \approx 10$, \hen{} should be placed at a distance of at least
10\,\kpc.  Since it is at a small galactic longitude in the plane of the
galaxy, one would expect an interstellar visual extinction of the order of
10\textsuperscript{m}, which is clearly not the case \citep[$\Av \lesssim 3$,
see][Table 2]{Elia04}.  Therefore, we deem that \hen{} does not belong to the sgB[e]
class.

\subsection{The Herbig Ae/Be hypothesis}

We were able to model \hen{} with a parametric disc model featuring the
properties often encountered in Herbig Ae/Be stars: an inner rim at the
sublimation temperature accounting for the 1-$5\,\micron$ flux and the
remaining, cooler parts of the disc that dominate the 10-$100\,\micron$ SED.
However, the properties of the YSO in this scenario are rather unusual and
contradictory.

First, the disc features a large ($30\,\AU$ in radius), almost free-of-matter
(``wide gap scenario'') inner gap, whereas it undergoes a short-lived phase of
unusually high mass accretion: the mass flow is $\Mdot \sim
4\text{--}250 \times 10^{-4} \Msun/\yr$, depending on accurate distance -- we
``liberally'' assume from 1 to 4\,\kpc.  It lays above the typical one in FU~Ori
outbursts \citep{HK96}.  If such an accretion rate is correct, we conversely
expect the inner gap to be filled with optically thick gas (``small gap
scenario'').  

If we further follow the common HAe/Be picture, the H$\alpha$ line originates
from rotating material, so that the peak-to-peak of $\approx 70$\,km/s
separation measured by \citet{Zickgraf03} implies material rotating at $\approx
35/\sin i$ km/s around the star.  With a typical young early B star mass of
10-15\,\Msun{}, the material must be located within 5\,\AU{} from the star,
once again pointing towards the ``small gap scenario''.  

Lastly, the P~Cygni profiles characteristic of HAe/Be stars are absent in the
spectrum of \hen{}.

Considering the previous points, the Herbig Ae/Be disc scenario is therefore
unlikely, though we cannot completely rule it out.  An alternative explanation
is that the mid and far-IR excess is not to be imputed to a strongly accreting
disc only, but to the presence of an envelope or wind in addition to a
disc.  This scenario would allow smaller accretion rates and be consistent with
absent P~Cygni profiles and the absence of gas in the inner gap of the disc; it
would not require that we observe an unlikely outburst.  However, this
alternative scenario requires more elaborate modelling that we have not
performed yet.

\subsection{The protoplanetary nebula hypothesis}

\begin{figure}
  \includegraphics[width=\linewidth]{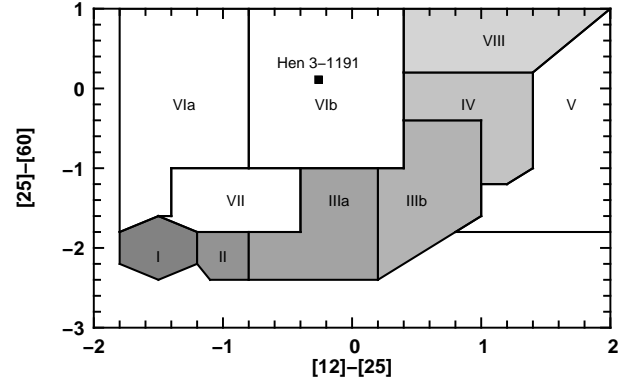}
  \caption{Position of Hen 3-1191 in the IRAS colour-colour diagram and
  comparison with the classification by \citet{Veen88}.  Shaded regions
  indicate the evolutionary track of post-AGB objects.  (Zero magnitude
  at 1\,Jy in all bands.)}
  \label{fig:evol-PPN}
\end{figure}
The large-scale (4'') bipolar nebula observed in the optical by
\citet{DeWinter94} as well as the similarity of its spectrum with that
of the Luminous Blue Variable $\eta$~Car and those of some confirmed 
proto-PNs \citep{Bertre89} are strong arguments in support of this 
hypothesis. We note, however, that the IRAS colours of Hen 3-1191 do not fit 
well within characteristics of post-AGB objects given by 
\citet{Kwok93,Veen88}, as shown in Fig.~\ref{fig:evol-PPN}
(region VIa, mostly made of unidentified Be or variable stars).  

The absence of high excitation lines \citep[][e.g. {[}\ion{O}{iii}{]} 
{[}\ion{N}{iii}{]}]{Landaberry01} as well as the hot ($\approx$ 1000\,K) mid-IR
temperature hints towards a dense, compact, and young post-AGB
object. Moreover, the linear size of the CS material obtained with our MIDI
measurements is
\begin{equation}
  r \approx 30 \left(\frac{d}{1\,\kpc}\right) \ \AU,
\end{equation}
where $d$ is the distance to the object.  With an upper limit
around 5\,kpc from the relatively low extinction, this places
the hot material at a distance of at most 150\,AU from the central
star, which indicates that the mass loss is recent or still
ongoing.

There is strong evidence that proto-PNs feature either a disc-like environment
or a highly asymmetrical shell with a wide polar cone containing optically thin
or moderately optically thick material \citep{Kwok93,Zickgraf03}.  Our
model of Sect.~\ref{sec:disk} yields an equatorial mass loss rate $\Mdot
\gtrsim 4\times10^{-5}\,\Msun/\yr$ using $\Mstar \le 8\,\Msun$ \citep[maximum
mass of proto-PNs after][]{Kastner05p}, $\Rmax \le 2500\,\AU$ \citep[value
observed in another object by][a likely upper limit for a young
proto-PN]{Kwok00}, and a conservative distance estimate ($d \ge 1\,\kpc$).
This is of the order of magnitude of the total (i.e. equatorial and bipolar)
mass loss rates modelled from proto-PN observations by \citet{Hrivnak05}.
However, the values specifically modelled in excretion discs around other types
of Be stars differ by several orders of magnitudes \citep[from $\ll 10^{-9}$ to
$> 2.5 \times 10^{-5}\,\Msun/\yr$ cited by][]{Lee91,Okazaki01,Kraus07}, so that
we cannot draw a conclusion concerning the validity of our mass loss rate.

\subsection{The symbiotic star hypothesis}

Our binary model, with a small separation of at least $\approx
18\,\AU$ per kpc of distance does not fit well with the accepted
picture of symbiotic stars \citep{Kenyon84}: a hot companion accreting material
from a very close late-type companion.  In addition, the characteristic size of
the CS dust in most symbiotic stars -- except those with a Mira star -- is
reported to be the order of 1\,\AU{} by \citet{Leedjarv06}, while the uniform
disc diameter of \hen{} is $\approx$ 30\,\AU{} per kpc of distance.

Furthermore, \hen{} does not show evidence of the presence of a late-type
companion \citep[absence of TiO lines according to][]{Landaberry01} and no high
ionisation lines ([\ion{O}{iii}] \& [\ion{N}{iii}], same authors), that are
standard characteristic of dusty symbiotic stars.  Furthermore this scenario
requires a large variability ($> 1$\,mag in the J band) in the near-IR as
measured in such symbiotic stars by \citet{Feast77}.  On the contrary,
\citet{Bertre89} measured little variability ($< 0.1$\,mag over the near-IR).

We therefore conclude that \hen{} is not a symB[e].


\section{Conclusion}

We observed and resolved the innermost parts of the CS matter around the B[e]
star \hen{} using MIDI on the VLTI, yielding a uniform disc
diameter of 24 to 36\,mas over the $N$ band.  Together with the SED, these
observations can be explained with a CS disc featuring a wide inner gap and an
unusually high accretion or excretion, whereas an envelope is not able to
account for them.  A binary system made of two unresolved IR sources with
different SED slopes in the IR may also account for the MIDI visibilities; we
did not investigate this hypothesis in detail, since the available data only
puts very loose constraints on such a system.

However, these observations do not allow us to give a definitive answer about
the nature of the object. The supergiant and symbiotic star scenarios are
extremely unlikely, so that \hen{} is most likely a young, compact proto-PN or
a Herbig Be star. However, in the Herbig HAe/Be scenario, the model
parameters reveal an atypical nature, though the additional presence
of an envelope or wind could explain part of the strong mid-IR flux.

Additional observations are required to determine the true nature of \hen{}.
Wide field imaging may discover a nearby star forming region;
diffraction-limited observations in the optical may also unveil bipolar nebular
structure characteristic of PNs; high resolution spectroscopy may also be a key
to the chemistry of the object, thus its evolutionary stage.  Additional
observations at the VLTI, in the near and mid-IR with AMBER and MIDI will also
be of use to obtain better constraints on the geometry of the CS matter and to
confirm or invalidate the accretion disc scenario with or without envelope, or
the binarity.

\begin{acknowledgements}
  This work has made use of NASA's Astrophysics Data System.  It also
  made use of the SIMBAD database and Vizier catalogue access tool,
  both operated at CDS, Strasbourg, France.  Most computations and
  graphics have been made with the free software Yorick developed
  by David H. Munro.  We thank the anonymous referee for a constructive
  report which helped to improve the quality of this paper.
\end{acknowledgements}


\bibliographystyle{aa}
\bibliography{biblio}

\end{document}